# The Dilemma of Standardizing Indoor Photovoltaic Characterisation: Embracing Diversity for Powering the IoT

<u>Zacharie Jehl Li-Kao</u>[1], Kunal J. Tiwari[1], Sergio Giraldo, Marcel Placidi[1], Yuancai Gong[1], Arindam Basak[2], Taizo Kobayashi[3], Jon Major[4], Edgardo Saucedo[1]

1. Polytechnic University of Catalonia, Spain
2. Kalinga Institute of Industrial Technology, India
3. Ritsumeikan University, Japan
4. University of Liverpool, UK

*Abstract:*

*In this viewpoint contribution, we argue that the emerging landscape of indoor photovoltaics poses unique challenges that transcend the capabilities of a singular standard, unlike what the community has become accustomed with the success of the AM1.x standard for outdoor application. We aim at illustrating the pitfalls associated with a one-size-fits-all approach to standardisation, emphasising the necessity for a concerted and nuanced methodology tailored to the complexities of indoor energy utilisation, and particularly in the context of the various needs of the Internet of Things. Acknowledging the inherent variability in indoor illumination conditions, and using simple numerical modelling and real-life examples to illustrate how it influences the output of indoor cells, we advocate for a shift from conventional standards to comprehensive guidelines that will better accommodate and evaluate the diverse interplays between photovoltaic device, internet of things sensors, and illumination sources. Our proposed methodology is not merely a set of rules but a strategic framework for the community to build upon, inviting researchers and industry stakeholders to collaborate and establish a unified foundation for assessing the performance of photovoltaic devices indoors. By fostering a collective approach and steering clear of rigid standards, this viewpoint lays the groundwork for future studies to better assess the performance and usability of indoor photovoltaics, thus ensuring innovation, adaptability, and reliable analyses in this very fast evolving and increasingly relevant field.*

Standardisation within the scientific community has become a cornerstone of the scientific method for both clarity reasons and comparison purposes, as a powerful tool to navigate in the maze of cutting-edge technological advances. In the domain of indoor photovoltaics (IPV), a standardised metric similar to AM1.5g used for outdoor applications is an attractive proposition, promising a more streamlined evaluation of different technologies while communities from organic PV, silicon, to inorganic thin film and perovskite sectors are trying to convince the world that their approach is the most suited for this emerging application. The perspective of standardisation is undeniably strong as such a benchmark would streamline comparisons. However, as one will evaluate the distinct needs of indoor photovoltaics, we ought to make a critical observation: the pursuit of a universal standard for indoor environments is a paradox, and potentially a pointless endeavour at best and dangerous pitfall at worst.

**Divergent Goals: Differentiating Outdoor and Indoor Photovoltaics**

The essence of outdoor and indoor PV systems diverges at a fundamental level. Outdoor PV, which has finally become a well-established solution for worldwide power generation, aims for a ubiquitous energy production regardless of the final application; whether for consumer electronics, electric vehicles, or home and public lighting, the efficiency under AM1.5g is the universal evaluation standard, and it is indeed an excellent metric despite geographical and environmental variations. On the other hand, indoor PV solar cells are to be engineered to power discrete appliances, primarily the emergent low-powered devices forming the Internet of Things (IoT). This field brings a variety of new challenges, ranging from varying power ratings and voltage thresholds for specific appliances to the optimal device position within a room due to the nature of the targeted application.

The variation in illumination sources in indoor environments brings another significant challenge when aiming for a standard. IoT devices are expected to be placed under lights, in most cases LED bulbs, with diverse temperature ratings and consequently with different emission spectra. For instance, IoT sensors used to monitor environmental conditions such as temperature/humidity, or air quality, are often installed in comparatively elevated positions for a broad coverage to improve data collection and quality. They might therefore be strategically placed near a given light source to maximize the power received by the indoor solar cells. On the other hand, sensors which are designed to detect moisture or leaks are often placed near possible sources of water like sinks, basements, or pipes, are expected to be typically located closer to the ground and therefore, often away from a direct sources of room illumination. This discrepancy in light positioning presents a combination of distinct possibilities that no single figure of merit can reliably navigate.

Finally, unlike the sun's relatively consistent nature within the majority of habitable outdoor settings, indoor environments have a wide array of illumination conditions due to factors such the proximity to a window, day/night cycles while the light stays on (very common in office buildings), dynamic environmental perturbations (someone temporarily standing in front of the device), and the diversity of light types used in different settings; for example, warm lights being preferred at home while cold/neutral lights are more common in professional environments. To illustrate this variability, we conducted simple measurements within a typical office space of our laboratory, illuminated by 5000K white LED sources (Figure 1). Four distinct locations are selected (labelled 1, 2,3 and 4) representing specific possible placements of an IoT device within the room: close to both the light source and a window (1), far from a light source but close to the window (2), close to a light source and distant from a window (3), and far from both the light source and a window (4). As shown in the corresponding table, a variation of the incident power density by two orders of magnitude is measured, from 1.9 mW.cm$^{-2}$ down to 0.07 mW.cm$^{-2}$. This variability does not even account for the day/night cycle, though indoor energy generation for IoT applications is anticipated to continue regardless of external lighting conditions, unlike outdoor PV where the intermittency in production is deemed acceptable and can be tackled through solutions such as energy storage and/or diversification.

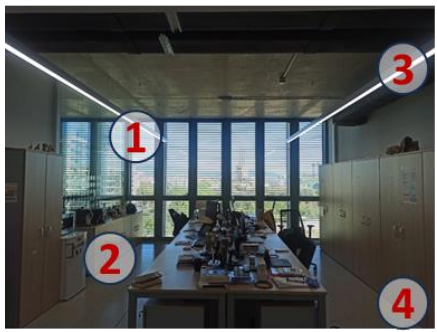

| | | |
|---|---|---|
| 1 | Power (mW.cm$^{-2}$) | 1.9 |
| | Luminance (lx) | ~9000 |
| 2 | Power (mW.cm$^{-2}$) | 0.09 |
| | Luminance (lx) | ~320 |
| 3 | Power (mW.cm$^{-2}$) | 1.55 |
| | Luminance (lx) | ~7800 |
| 4 | Power (mW.cm$^{-2}$) | 0.07 |
| | Luminance (lx) | ~280 |

*Figure 1. Illustration of the variability in indoor lighting in a typical office space.*

To exemplify the influence of the light source variability, we model a simple p-n homojunction operating in the radiative limit using SCAPS with three bandgaps considered: 1.7eV, 1.8eV, and 1.9eV, each within the range considered ideal for capturing indoor LED light [1]. This simple model, not deemed quantitative, is presented in order to show how the nature of the source illumination can drastically influence performance. The corresponding efficiencies are presented in surface plots Figure 2 wherein both the temperature of the light source and the incident power density are being varied, from 2700K to 6000K and from 0.1 mW.cm$^{-2}$ to 2mW.cm$^{-2}$ respectively.

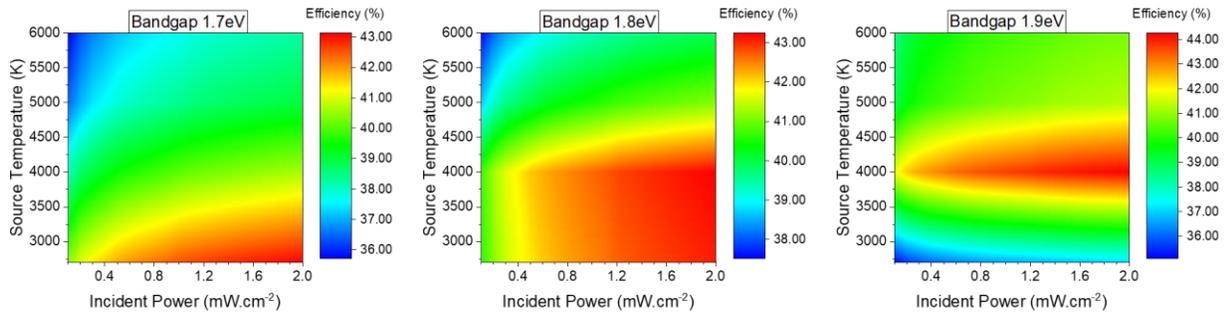

*Figure 2.Modelling of the performance of a simple pn junction in the radiative limit under variable LED indoor illumination (source temperature and incident power).*

Those figures clearly demonstrate the existence of an optimal source temperature for each bandgap, trivially understood by the different spectral matching between the absorbers and the sources. Significant variations in the efficiencies of nearly 25% in relative units are observed for all considered absorber bandgaps. Moreover, one can see that the changes in power density also markedly influences the resulting cell performance, an effect which in a realistic device can be further enhanced by non-radiative recombination. As a practical demonstration under realistic conditions, Figure 3 presents a quantitative model featuring Silicon ($E_g = 1.12eV$), CdTe ($E_g = 1.5eV$), and Sb$_2$S$_3$ ($E_g = 1.85eV$) solar cells as reference materials. Unlike the idealised conditions in Figure 2, this model incorporates a realistic set of defects and recombination processes. Two key observations emerge from this analysis. Firstly, it becomes evident that targeting bandgaps within the range of 1.7eV to 2eV is crucial when designing indoor PV devices. Notably, Sb$_2$S$_3$, which starts with a modest 6.8% AM1.5 efficiency outdoors, shows a remarkable nearly threefold increase in conversion efficiency to levels exceeding 18% for source temperatures below 4000K. Secondly, devices with bandgaps closer to the optimum value demonstrate greater resilience to variations in source temperature, a notion which will be further discussed in this article. The fact that a specific bandgap inherently aligns more favourably with certain light sources is a departure from the outdoor characterisation paradigm. Indeed, the detailed balance limit demonstrates a plateau in performance for bandgap values between 1.1eV and 1.5eV. No such plateau exists even in the narrow bandgap range from 1.7eV to 1.9eV. This important difference serves as a fundamental justification for the premises outlined in this article.

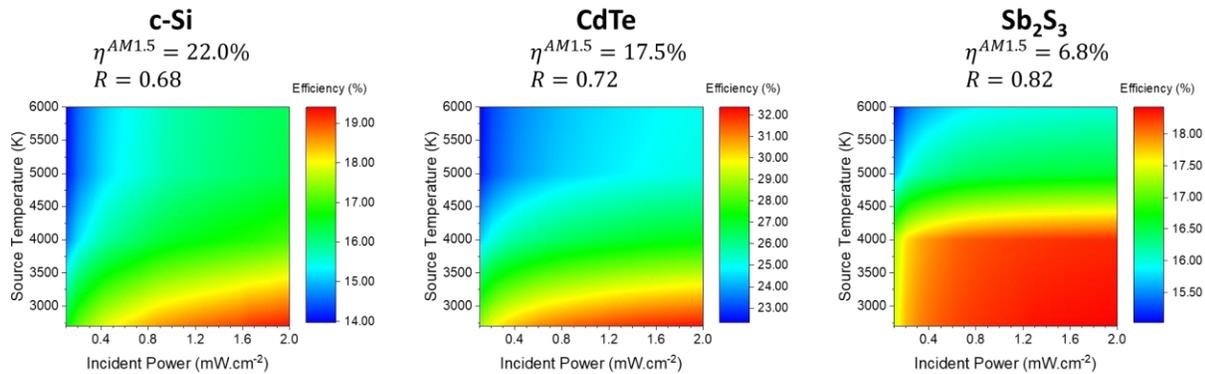

*Figure 3. Modelling of the performance of three reference technologies under variable LED indoor illumination (source temperature and incident power).*

Those models, while simple, illustrate how a single standard would inadequately encompass the inherent variability of indoor illumination, and how different technologies may answer different needs. Rather than perceiving this variability in indoor conditions as a limitation, we advocate to welcome it as an opportunity to establish a robust methodology highlighting the importance of the scientific method for a reliable analysis in the landscape of indoor photovoltaics.

**Rethinking our approach by Embracing diversity in Characterisation and proposing a paradigm shift for Indoor PV**

The perspective and attractivity of establishing a standard for indoor PV is undeniable, considering the scientific success it promises to its authors and the necessity to streamline comparisons across the large variety of technologies, which has been a staple for the PV community since its inception. Beyond our belief that the variation in the incident power density and large variety of light types and IoT devices are a problem that no single standard can solve, one should keep in mind that attempting to create a single standard for indoor would inevitably favour certain technologies over others, much more so than for outdoor PV. Thus, veering the research in indoor PV towards a specific lighting scenario and possibly motivated by financial interests in a market bound to grow exponentially within the next decade, rather than a really broad and holistic adaptability to the diverse indoor conditions.

These limitations in the quest for standardisation, primarily originating from the intricate diversity of indoor environments and the inadequate metrics and reflexes inherited from outdoor PV habits, will necessitate a radical shift in our approach to accurately evaluate indoor photovoltaics for IoT applications. Instead of what we believe to be the futile pursuit of a one size fits all standard, a more scientifically relevant and pragmatic approach revolves in our opinion around understanding the interplay between the indoor PV cells and their expected application in the context of IoT. Rather than a standard, we propose in the viewpoint article a method, and guidelines for the community to build upon.

We believe that as a first step, it would be necessary to exhaustively map and categorize IoT devices according to characteristics deemed relevant to power generation, to their footprint and their final application; that is, their distinct power and voltage requirements, size, and the illumination conditions in which those devices are expected to be operating. This necessitates a close interaction between the PV community and several of the relevant IoT stakeholders. Large PV consortia, such as the recently formed RENEWPV European initiative (https://renewpv.eu), offer unique perspectives and a solid framework in that regard and could be instrumental in fostering this sort of interaction with the IoT community.

Secondly, on the PV side, we believe that the introduction of a resilience factor in our indoor PV characterisation routines is a strong necessity to evaluate the versatility of a solar cell in indoor conditions. For simplicity reasons, it is recommended that this resilience factor would be defined for a given illumination temperature T. This new figure of merit, measured as the ratio of output power between the expected lowest and highest indoor incident power conditions characterised using an experimental framework where a range of power densities are considered, offers a clear and direct technology-agnostic assessment of a PV cell's adaptability across varying lighting scenarios. The resilience factor is defined for a specific source temperature as

$$R^T = \frac{\eta_{0.1}^T}{\eta_2^T}$$

Where $R^T$ is the dimensionless resilience factor for a source temperature T, $\eta_{0.1}^T$ is the measured efficiency for T with an incident power of 0.1mW.cm$^{-2}$, and $\eta_2^T$ is the measured efficiency for T with an incident power of 2mW.cm$^{-2}$. The values 0.1 and 2 are taken after our own measurements from Figure 1 and while we do believe that those values properly encompass the overwhelming majority of possible injection conditions, those should be taken as guidelines until a proper consensus emerges within the community. Moreover, and in the case of more exhaustive characterisations where several temperatures are considered and a surface plot similar to the model from Figure 3 is experimentally obtained, the resilience factor can be generalised as R. We report as an example Figure 3 the R factor for c-Si, CdTe and Sb$_2$S$_3$. The latter, despite the higher defect density which would normally make it less resilient to low light intensities, reaches the highest R value at 0.82

A resilience factor which is close to 1 will denote versatility, allowing the technology to function proficiently and consistently across a broader range of indoor conditions, while a lower ratio implies a solar cells more suited for specific, more tailored lighting scenarios with higher incident power densities. Considering the objective of replicating diverse indoor lighting scenarios for more accurate experimentations, as modelled Figure 2, the recent emergence of LED-based solar simulators is bound to become an essential tool for research groups aiming at evaluating their PV technology in various indoor conditions as they possess the ability to individually adjust LEDs and mimic a spectrum of indoor lighting conditions thus permitting a more accurate and relevant evaluation of PV cell performance across varied environments.

**Establishing Practical Standards: The importance of the Footprint**

We established that a reinforced relationship between the PV sector and IoT stakeholders is imperative for establishing new standards tailored to the requirements of IoT devices, rather than catering for a specific PV technology; in a nutshell, to design indoor cells tailored to the devices being powered rather than the other way around. We aim at synthesising the two previously discussed points; that is categorizing IoT devices based on their energy requirements and anticipated locations within indoor spaces, and evaluating the resilience of our indoor cells in different photon injection scenarios, to more accurately communicate on a technology's performance. In essence, it means reporting a tangible figure of merit: the required surface area of the solar cell to effectively power a specific class of IoT system. In this envisaged framework, the true measure of success is now linked to the minimisation of the surface area, rather than the raw performance for a given scenario. The ultimate goal transcends the concept of efficiency as defined for outdoor applications, emphasising instead the optimisation of space utilisation, which we believe is a crucial aspect in the practical integration of photovoltaic technologies within the fast-evolving landscape of indoor IoT applications where miniaturisation remains the paramount.

**Conclusion: Complexity for Better Applicability**

The question of standardizing indoor photovoltaic characterisation is a complex issue, owing to the diversity both in illumination conditions and power requirements of the targeted devices. The pursuit of a universal standard, similar to AM1.5g for outdoor PV, is we believe futile when considering the variety of indoor environments and devices. The diverse needs of IoT devices in different indoor lighting scenarios require for a paradigm shift, driving us away from a single benchmark and toward a more adaptable and scientific accurate approach. The divergence in end goals between outdoor and indoor photovoltaics demonstrates the necessity for a tailored and application-oriented evaluation method.

By accepting and even embracing this complexity, a novel approach can arise, shifting the focus towards the understanding and categorizing IoT devices according to their specific power requirements and expected positions within indoor settings as compared to the light source. The introduction of a resilience factor offers a clear assessment of a PV cell's adaptability to various lighting scenarios, while modern tools such LED-based solar simulators can provide the means to replicate and to evaluate diverse indoor conditions accurately across different research groups.

The interaction of IoT and PV sectors in establishing practical standards aligns with the ultimate goal of powering IoT devices with minimal surface area. This collaboration should foster benchmarks that are not only pragmatic in terms of experimental feasibility but also properly reflect the diverse landscape of IoT devices in indoor settings.

In conclusion, the necessity for a nuanced and adaptable approach cannot be overstated, rather than a rigid standard which would kneecap innovation and limit the exploration of technologies catering to the multifaceted needs of IoT devices in indoor environments. Embracing the complexity inherent in these settings will lead to more accurate and relevant standards, facilitating advancements in indoor PV technologies and their integration into the expanding domain of IoT, while limiting the possible interference of private interests not necessarily motivated by scientific probity.